\documentclass[aps,pra,onecolumn,preprint,preprintnumbers,groupaddress,amsmath,amssymb,usenatbib,
nofootinbib,showkeys,showpacs,altaffilletter,11pt]{revtex4}
\usepackage{graphicx}
\usepackage{dcolumn}
\usepackage{bm}
\usepackage{latexsym}
\usepackage{amsfonts}
\usepackage{amssymb}
\usepackage{amsmath}
\usepackage{subfigure}
\usepackage{mathrsfs}
\usepackage{tabu}
\usepackage[colorlinks]{hyperref}

\begin{document}

\title{Stability and interacting $f(T,\mathcal{T})$ gravity with modified Chaplygin gas}

\author{T. Mirzaei Rezaei}
\email{st.t.mirzaei@iauamol.ac.ir}
\author{Alireza Amani}
\email{a.r.amani@iauamol.ac.ir}
\affiliation{\centerline{Faculty of Sciences, Department of Physics, Ayatollah Amoli Branch, Islamic Azad University,} \\Amol, Mazandaran, Iran}

\date{\today}

\begin{abstract}
In this paper, the model of interaction is studied between $f(T,\mathcal{T})$ gravity and modified Chaplygin gas in FRW-flat metric. We obtain the Friedmann equations in the framework of teleparallel gravity by vierbein field. We consider that Universe dominates by components of cold matter, dark energy and modified Chaplygin gas. In what follows we separately write the corresponding continuity equations for components of Universe. Also, dark energy EoS and effective EoS are obtained with respect to redshift, thereafter the corresponding cosmological parameters are plotted in terms of redshift, thereinafter the accelerated expansion of the Universe is investigated. Finally, the stability of the model is discussed in phase plane analysis.
\end{abstract}
\maketitle

\section{Introductionn}\label{s1}

The first time expansion of Universe discovered in type Ia supernova (SNe Ia) by Riess et al \cite{Riess_1998}, and then Perlmutter et al \cite{Perlmutter_1999} redemonstrated this issue by $42$ supernovae. Also accelerated expansion of Universe reconfirmed by cosmic microwave background (CMB) \cite{Bennett_2003} and large scale structure (LSS) \cite{Tegmark_2004}. Thus, the result of these literature confirms the existence of a mysterious energy called dark energy. One of the biggest challenges of modern theoretical physics is understanding the nature of dark energy. In fact, dark energy is a hypothetical form of energy in space, and constitutes three--quarters of the total energy of the Universe. Dark energy requires strong negative pressure to explain the observed accelerated expansion of the Universe. By using the Einstein field equation, the accelerated expansion is described by a small positive cosmological constant. Also, the discovery illustrates that geometry of the Universe is very close to flat space-time \cite{Weinberg-1989}. In order to describe dark energy model, numerous studies have been performed in isotropic space-time such as scalar fields (quintessence, phantom,
tachyon and etc.) \cite{Caldwell-2002, Amani-2011, Sadeghi1-2009, Setare-2009, Setare1-2009, Battye-2016, Li-2012, Khurshudyan-2014, Khurshudyan-2015, Sadeghi-2013, Sadeghi-2014, Pourhassan-2014, MKhurshudyan11-2015, Banijamali1-2016,Banijamali2-2012,JSadeghi-2015, MKhurshudyan1-2015, Behrouz-2017}, modified gravity models \cite{Amani1-2015, Iorio-2016, Faraoni-2016, Khurshudyan1-2014, Banijamali-2016, Sadeghi1-2016, Sadeghi2-2016, Banijamali3-2012}, holographic models \cite{Wei-2009, Amani1-2011, Amani-2015, Li-2004, Campo-2011, Hu-2015, Fayaz-2015, Saadat1-2013, Banijamali3-2011}, interacting models \cite{Amani-2013, Amani-2014, Naji-2014, KhurshudyanJ-2014, KhurshudyanB-2014, Morais-2017, Zhang1-2017, Bouhmadi1-2016, KhurshudyanJ1-2015, KhurshudyanJ2-2014, SadeghiKhurshudyanJ-2014, SadeghiKhurshudyanJ1-2014}, bouncing model \cite{Sadeghi-2010, Sadeghi-2009, Amani-2016, Singh-2016} and braneworld models \cite{Sahni-2003, Setare-2008, Brito-2015}.

Now we intend to describe dark energy model with modified gravity theory. This issue studied in numerous literatures by topics of $f(R)$ gravity \cite{Nojiri_2006, Nojiri_2007}, $f(T)$ gravity \cite{Linder_2010, Myrzakulov_2011}, $f(\mathcal{T})$ gravity \cite{Li_2011}, $f(R,T)$ gravity \cite{Myrzakulov_2012, Amani1-2015, Amani-2015}, $f(R,\mathcal{T})$ gravity \cite{Harko_2011} and $f(G)$ gravity \cite{Nojiri_2005}, in which $R$, $T$, $\mathcal{T}$ and $G$ are Ricci scalar curvature, torsion scalar, trace of the matter energy-momentum tensor and the Gauss-Bonnet term, respectively. We note that cosmological solutions of these models can provide an alternative explanation for the accelerated expansion of Universe.

In general relativity, the Einstein-Hilbert action includes the curvature term that describes gravity, but in teleparallel gravity, the corresponding action includes torsion term that first proposed by Einstein \cite{Einstein_1928}. He has been able to introduce the mathematical structure of distant parallelism by a tetrad or vierbein field for unification of electromagnetism and gravity. This issue is caused to replace the Levi-Civita connection in the framework of general relativity with the Weitzenb\"{o}ck connection in teleparallelism \cite{Weitzenbock_1923, Bengochea_2009}. This means that the $R$ in the general relativity changes to $T$ in teleparallelism. Its modified form has been done by changing $T$ to $f(T)$ with an arbitrary function in the teleparallelism action so-called $f(T)$ gravity theory.

A model recently proposed by subject of $f(T,\mathcal{T})$ gravity, that one has some benefits in comparison with other models of modified gravity theory \cite{Harko_2014}. One of its benefits is much simpler calculations of numerical solutions, and also is compatible with recent observational data to describe accelerated expansion of the Universe. Since the $f(T,\mathcal{T})$ gravity can be a good candidate for a source of dark energy, then one is a good alternative for the standard gravity model. Hence these are a good motivation for the present job.

One of the other interesting models for the description of dark energy is Chaplygin gas, so one is introduced as a fluid with negative pressure. Chaplygin gas model extended to generalized Chaplygin gas and modified Chaplygin gas, in which the modified Chaplygin gas model is the combination of barotropic model and Chaplygin gas model. The advantage of the modified Chaplygin gas model is consistent with observational data, and also unify dark matter and dark energy \cite{Kamenshchik_2001, Amani1-2015, Amani-2013, Amani-2014}. Also, other models of Chaplygin gas such as viscous Chaplygin gas, cosmic Chaplygin gas and extended Chaplygin gas studied by Refs. \cite{Marttens-2017, Jawad-2017, Kahya-2015, Avelino-2014, Pourhassan-2013, Kahya-2014, Pourhassan-2016, KahyabB-2015, SaadatA-2014, NajiJ-2014, SaadatB-2013, SaadatC-2013, SadeghiB-2016, SaadatyaB-2013, SadeghiaB-2014, SaadataB-2014, KhodamB-2016, SaadatmB-2013, SaadatmBB-2013, PourhassanBB-2014, PourhassanmBB-2014, KahyaB-2015,PourhassanatmBB-2016}. But in this paper we use from a model not so complicated as the modified Chaplygin gas model.

Now, we would like to construct a new model based on interacting $f(T,\mathcal{T})$ gravity with the modified Chaplygin gas. The corresponding model may be a development of several previous jobs that also corresponds with observational data.

This paper is organized as the following:

In Sec. \ref{s2}, we review the general form of $f(T,\mathcal{T})$ gravity in the FRW metric. In Sec. \ref{s3}, we consider the interaction between $f(T,\mathcal{T})$ gravity model and the modified Chaplygin gas, and also obtain the Friedmann equations and equation of state (EoS). In Sec. \ref{s4}, we investigate the stability of our model by phase plane analysis. Finally, in Sec. \ref{s5} we will give result and conclusion for our model.


\section{$f(T,\mathcal{T})$ gravity model}\label{s2}
We start with a novel theory of modified gravity in which there is an arbitrary function of torsion scalar $T$ and trace of the matter energy-momentum tensor $\mathcal{T}$. In that case, we can write the corresponding action by
\begin{equation}\label{action1}
  S=\int e \left[\frac{T+f(T,\mathcal{T})}{2 k^2}+\mathcal{L}_m\right] d^4x,
\end{equation}
where $e=det\left( e^i_{\,\,\mu}\right)$, and $\mathcal{L}_m$ is the matter Lagrangian density. It should be mention that vierbein field $e^i_{\,\,\mu}$ to be related with metric tensor by relationship $g_{\mu \nu}=\eta_{\mu \nu} e^i_{\,\,\mu} e^i_{\,\,\nu}$ in which \({{\eta }_{ij}}=diag(-1,+1,+1,+1)\).

In this paper, Universe is considered  as the homogenous and isotropic in the FRW metric in the following form
\begin{equation}\label{ds2}
d{{s}^{2}}=d{{t}^{2}}+{{a}^{2}}(t)(d{{x}^{2}}+d{{y}^{2}}+d{{z}^{2}}),
\end{equation}
where $a(t)$ is scale factor. The vierbein field can be written in accordance with metric \eqref{ds2} as \({{e}^{i}}_{\mu }={{e}^{\mu }}_{i}=diag(1,a,a,a)\). We note that the Levi-Civita connection is as
\begin{equation}\label{levi1}
 \Gamma _{\mu \nu }^{\rho }=\frac{1}{2}g_{{}}^{\rho \sigma }\left( {{\partial }_{\nu }}{{g}_{\sigma \mu }}+{{\partial }_{\mu }}\,{{g}_{\sigma \nu }}\,-\,{{\partial }_{\sigma }}{{g}_{\mu \nu }} \right).
\end{equation}

Ricci Tensor $R_{\mu \nu}$, asymmetry tensor ${{S}_{\rho }}^{\mu \nu }$ and torsion tensor ${{T}^{\lambda }}_{\mu \nu }$ are written as follows, respectively
\begin{eqnarray}
 &R_{\mu \nu } = {{\partial }_{\lambda }}\Gamma _{\mu \nu }^{\lambda }-{{\partial }_{\mu }}\Gamma _{\lambda \nu }^{\lambda }+\Gamma _{\mu \nu }^{\lambda }\Gamma _{\rho \lambda }^{\rho }-\Gamma _{\nu \rho }^{\lambda }\Gamma _{\mu \lambda }^{\rho }, \label{RST-1}\\
 &{{s}_{\rho }}^{\mu \nu } =\frac{1}{2}\left( {{k}^{\mu \nu }}_{\rho }+\delta _{\rho }^{\mu }{{T}^{\alpha \nu }}_{\alpha }-\delta _{\rho }^{\nu }{{T}^{\alpha \mu }}_{\alpha } \right),\label{RST-2} \\
 &{{T}^{\lambda }}_{\mu \nu }={{\Gamma }^{\lambda }}_{\nu \mu }-{{\Gamma }^{\lambda }}_{\mu \nu }=e_{A}^{\rho }\left( {{\partial }_{\mu }}e_{\nu }^{A}-{{\partial }_{\nu }}e_{\mu }^{A} \right),\label{RST-3}
\end{eqnarray}
where
\begin{equation}\label{kmunu}
  {{k}^{\mu \nu }}_{\rho }=-\frac{1}{2}\left( {{T}^{\mu \nu }}_{\rho }+{{T}^{\nu \mu }}_{\rho }-{{T}_{\rho }}^{\mu \nu } \right).
\end{equation}

By varying the action \eqref{action1} with respect to vierbein field, we can obtain the field equation as follows:
\begin{eqnarray}\label{fri2}
\begin{aligned}
  {{S}_{\mu }}^{\nu \rho }{{f}_{TT}}\,{{\partial }_{\rho }}T+\left[ {{e}^{-1}}{{e}^{i}}_{\mu }{{\partial }_{\rho }}\left( e{{e}_{i}}^{\mu }{{S}_{\alpha }}^{\nu \lambda } \right)+{{T}^{\alpha }}_{\lambda \mu }{{S}_{\alpha }}^{\nu \lambda } \right]\left( 1+{{f}_{T}} \right)+\frac{1}{4}\delta _{\mu }^{\nu }\,T \\
  ={{S}_{\mu }}^{\nu \rho }{{f}_{TT}}\,{{\delta }_{\rho }}\mathcal{T} +{{f}_{\mathcal{T} }}\left( \frac{\Theta _{\mu }^{\nu }+\delta _{\mu }^{\nu }p}{2} \right)-\frac{1}{4}\delta _{\mu }^{\nu }\,f(\mathcal{T} )+\frac{{{k}^{2}}}{2}\Theta _{\mu }^{\nu }, \\
\end{aligned}
\end{eqnarray}
where indices denote the derivative with respect to the corresponding parameters, and symbol $\Theta _{\mu }^{\nu }$ is the Energy- momentum tensor of the matter.

Tension scalar and trace of energy-momentum tensor are obtained as the follows:
\begin{subequations}\label{ttheta}
\begin{eqnarray}
 &T={{T}^{\lambda }}_{\mu \nu }{{S}_{\lambda }}^{\mu \nu }=-6{{H}^{2}},\label{ttheta-1}\\
 &\Theta =\mathcal{T} =({{\rho }_{m}}-3{{p}_{m}}),\label{ttheta-2}
\end{eqnarray}
\end{subequations}
where $H$, $\rho_m$ and $p_m$ are the Hubble parameter, matter energy density and matter pressure, respectively.

\section{Interaction between $f(T,\mathcal{T})$ gravity and modified Chaplygin gas}\label{s3}

A recent cosmological model has been designed using a kind of perfect fluid (Chaplygin gas), that the universe is assumed to be filled with it. In this paper, the modified Chaplygin gas model was used and its mode equation can be expressed as follows:
\begin{equation}\label{pc1}
{{p}_{c}}=A{{\rho }_{c}}-\frac{B}{{{\rho }_{c}}^{\alpha }},
\end{equation}
where $A$, $\rho_c$ and $p_c$ are positive constant, energy density and pressure of modified Chaplygin gas, respectively and also we have $0 < \alpha < 1$.
In here we consider the Universe dominates by components of matter, dark energy and Chaplygin gas. In that case, the effective energy density and the effective pressure of Universe are written in terms of the aforesaid three components as
\begin{subequations}\label{rhoeffpeff}
\begin{eqnarray}
 {{\rho }_{eff}}={{\rho }_{m}}+{{\rho }_{DE}}+{{\rho }_{c}},\label{rhoeffpeff-1}\\
 {{p}_{eff}}={{p}_{m}}+{{p}_{DE}}+{{p}_{c}},\label{rhoeffpeff-2}
\end{eqnarray}
\end{subequations}
where indices $m$, $DE$ and $c$ are components of cold mater, dark energy and modified Chaplygin gas, respectively.

As we know, Universe consists a perfect fluid, then the corresponding continuity equation can be written in the following form
\begin{equation}\label{conti1}
  {{\dot{\rho }}_{eff}}+3H({{\rho }_{eff}}+{{p}_{eff}})=0.
\end{equation}

Since there is an energy flow between contents of Universe, thus we separately write the continuity equations of components of Universe as
\begin{subequations}\label{conti2}
\begin{eqnarray}
  &{{\dot{\rho }}_{m}}+3H({{\rho }_{m}}+{{p}_{m}})={Q}'-Q,\label{conti2-1}\\
  &{{\dot{\rho }}_{c}}+3H({{\rho }_{c}}+{{p}_{c}})=Q,\label{conti2-2}\\
  &{{\dot{\rho }}_{DE}}+3H({{\rho }_{DE}}+{{p}_{DE}})=-{Q}',\label{conti2-3}
\end{eqnarray}
\end{subequations}
where $Q$ and $Q'$ are interaction terms between the Universe components. Let us consider that there is an energy flow only between the components of dark energy and modified Chaplygin gas, in this way $Q=Q'$. Also, the amount of this interaction terms is considered as $Q={Q}'=3{{b}^{2}}H{{\rho }_{c}}$ in which $b^2$ is the coupling parameter or transfer strength. In this case, the solutions of Eqs. \eqref{conti2-1} and \eqref{conti2-2} are as follows:
\begin{equation}\label{hamilton}
 {{\rho }_{m}}={{\rho }_{{{m}_{0}}}}{{a}^{-3(1+{{\omega }_{m}})}},
\end{equation}
\begin{equation}\label{hamilton}
{{\rho }_{c}}={{\left[ \frac{B}{\eta }+{{c}_{0}}\,{{a}^{-3\eta (\alpha +1)}} \right]}^{\frac{1}{\alpha +1}}},
\end{equation}
where
\begin{equation}\label{hamilton}
 \eta =1+{{b}^{2}}+A.
\end{equation}

In this paper, in order to solve the Friedmann equations and to avoid its complexity, we consider a simple particular model of $f(T,\mathcal{T})$ gravity as a linear function of $T$ and $\mathcal{T}$ in the form \(f(T,\mathcal{T})=T+\gamma \, \mathcal{T} \). Therefore, Friedmann equation \eqref{fri2} is deduced as
\begin{subequations}\label{fri3}
\begin{eqnarray}
  &3{{H}^{2}}=-\frac{\gamma }{2}\mathcal{T} +2{{\rho }_{m}}+{{p}_{m}},\label{fri3-1}\\
  &-3{{H}^{2}}-2\dot{H}\,={{p}_{_{m}}}+\frac{\gamma }{2}\mathcal{T}.\label{fri3-2}
\end{eqnarray}
\end{subequations}

Now, the above equations can be written in Friedmann standard as follows.
\begin{subequations}\label{fri4}
\begin{eqnarray}
  &3{{H}^{2}}={{\rho }_{_{eff}}},\label{fri4-1}\\
  &-3{{H}^{2}}-2{\dot{H}}\,={{p}_{_{eff}}}.\label{fri4-2}
\end{eqnarray}
\end{subequations}

From Eqs. \eqref{fri3} and \eqref{fri4}, we can find the density energy and pressure of dark energy in the following form
\begin{subequations}\label{rhopde}
\begin{eqnarray}
  &{{\rho }_{DE}}=\left( 1+{{\omega }_{m}}-\frac{\gamma }{2}(1-3{{\omega }_{m}}) \right){{\rho }_{m}}-{{\rho }_{c}},\label{rhopde-1}\\
  &{{p}_{DE}}=\frac{\gamma }{2}\left( 1-3{{\omega }_{m}} \right){{\rho }_{m}}-{{p}_{c}},\label{rhopde-2}
\end{eqnarray}
\end{subequations}
where $\omega_m$ is the EoS of matter. Also, we define effective EoS of a perfect fluid, and EoS of dark energy as
\begin{equation}\label{omeeff}
 {{\omega }_{eff}}=\frac{{{p}_{eff}}}{{{\rho }_{eff}}}=-1-\frac{2\dot{H}\,}{3{{H}^{2}}},
\end{equation}
\begin{equation}\label{omede}
 {{\omega }_{DE}}=\frac{{{p}_{DE}}}{{{\rho }_{DE}}}.
\end{equation}

Therefore, Friedmann obtained equations are written based on the redshift parameter ($z=\frac{a_0}{a}-1$), in that case, the Eqs. \eqref{rhopde-1} and \eqref{rhopde-2} are rewritten by redshift parameter as
\begin{subequations}\label{rhopde1}
\begin{eqnarray}
  &{{\rho }_{DE}}={{\rho }_{{{m}_{0}}}}\left( 1+{{\omega }_{m}}-\frac{\gamma }{2}(1-3{{\omega }_{m}}) \right){{\left( \frac{{{a}_{0}}}{1+z} \right)}^{-3(1+{{\omega }_{m}})}}- {{\left[ \frac{B}{\eta }+{{c}_{0}}\,{{\left( \frac{{{a}_{0}}}{1+z} \right)}^{-3\eta (\alpha +1)}} \right]}^{\frac{1}{\alpha +1}}},\label{rhopde1-1}\\&
  \begin{aligned}
  &{{p}_{DE}}=\frac{\gamma }{2}\left( 1-3{{\omega }_{m}} \right){{\rho }_{{{m}_{0}}}}{{\left( \frac{{{a}_{0}}}{1+z} \right)}^{-3(1+{{\omega }_{m}})}}- A{{\left[ \frac{B}{\eta }+{{c}_{0}}\,{{\left( \frac{{{a}_{0}}}{1+z} \right)}^{-3\eta (\alpha +1)}} \right]}^{\frac{1}{\alpha +1}}}\\&
  +\frac{B}{{{\left[ \frac{B}{\eta }+{{c}_{0}}\,{{\left( \frac{{{a}_{0}}}{1+z} \right)}^{-3\eta (\alpha +1)}} \right]}^{\frac{\alpha }{\alpha +1}}}}.\label{rhopde1-2}
  \end{aligned}
\end{eqnarray}
\end{subequations}

\begin{figure}[h]
\begin{center}
\subfigure
{\includegraphics[scale=.3]{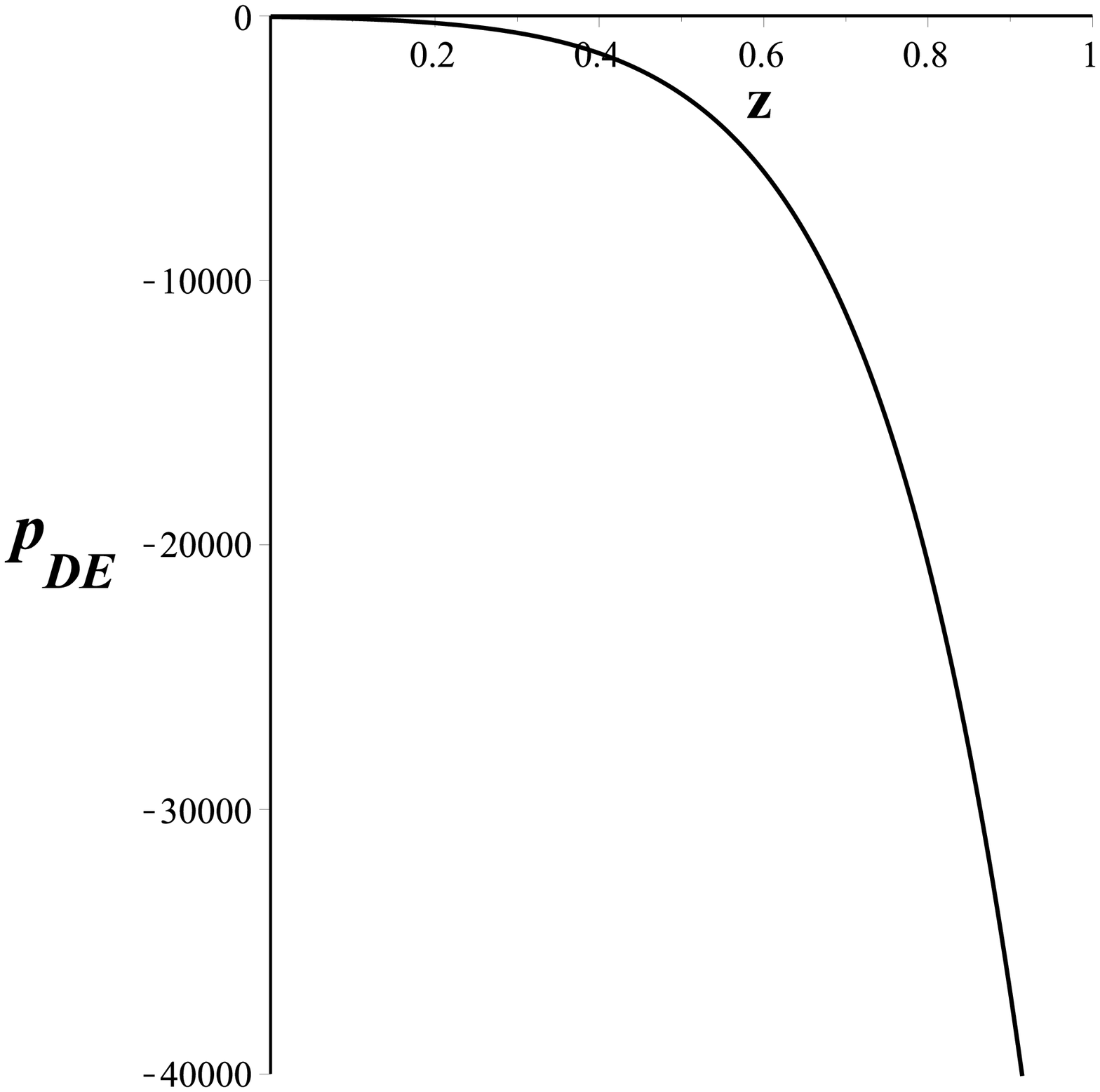}\label{fig1-1}}
\subfigure
{\includegraphics[scale=.3]{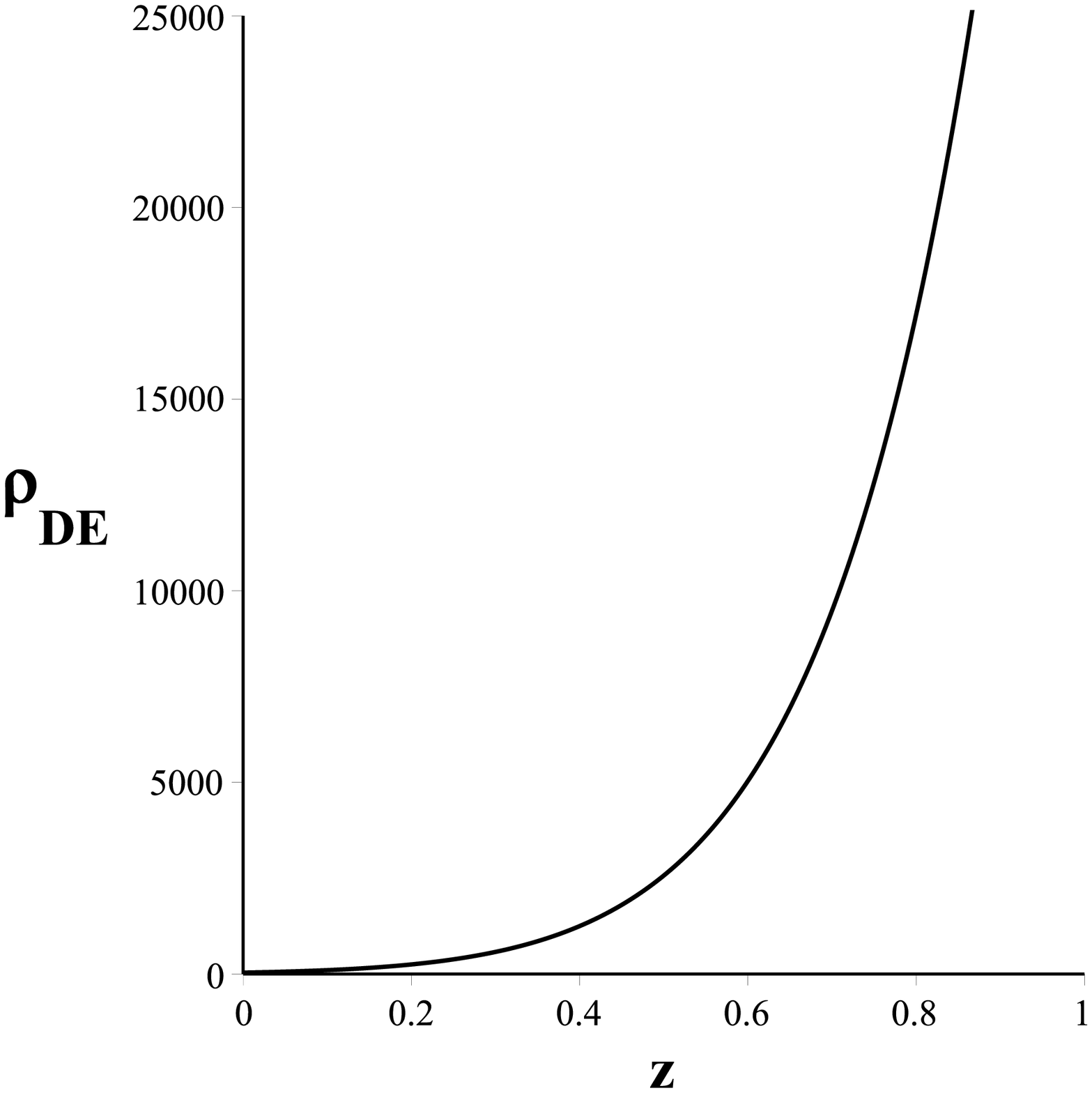}\label{fig1-2}}
\caption{The energy density and pressure of dark energy in terms of redshift for $A = 3$, $B = 5$, $b = 0.5$, $\gamma = 0.5$, $\rho_{m_0}=2$, $\omega_m = 2.5$, $a_0 = 1$ and $c_0 = 3.5$.}\label{fig1}
\end{center}
\end{figure}
By substituting Eqs. \eqref{rhopde1} into the Eq. \eqref{omede}, and plotting $\omega_{DE}$ versus $z$. In that case, we see the $\omega_{DE}$ at the present time $(z=0)$ equals to $-1.046$, which one indicates Universe expanding at an accelerating rate, and also the issue corresponds to work \cite{Shi_2011, Amanullah_2010, Kumar_2014}.
\begin{figure}[h]
\begin{center}
\subfigure
{\includegraphics[scale=.3]{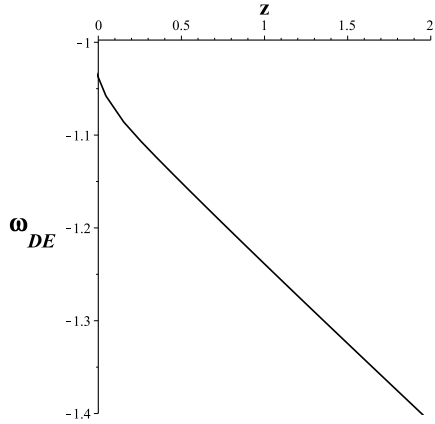}}
\caption{The dark energy EoS in terms of redshift for $A = 3$, $B = 5$, $b = 0.5$, $\gamma = 0.5$, $\rho_{m_0}=2$, $\omega_m = 2.5$, $a_0 = 1$ and $c_0 = 3.5$.}\label{fig2}
\end{center}
\end{figure}

By drawing figures of energy density $\rho_{DE}$ and pressure $p_{DE}$ of dark energy with respect to redshift (Fig. \ref{fig1}), we can see that the variations of $\rho_{DE}$ versus $z$ is positive, and the variations of $p_{DE}$ versus $z$ is negative, so that these issues confirm the accelerated expansion of the Universe. Also, the Figs. \ref{fig1} and \ref{fig2} show us that free parameters of the modified Chaplygin gas ($A$ and $B$) and the interacting term ($b$) play a very important role to confirm the accelerated expansion of the universe (the obtained results are consistent with \cite{KhurshudyanJ2-2014, SadeghiKhurshudyanJ-2014}).\\
\section{Stability Analysis}\label{s4}
In this section, we intend to investigate the stability conditions of the $f(T,\mathcal{T})$ gravity with modified Chaplygin gas by
the method of phase plane analysis. In that case, we introduce the following new variables as
\begin{equation}\label{newvar}
x=\,\frac{{{\rho }_{c}}}{3{{H}^{2}}},\qquad y=\frac{{{\rho }_{m}}}{3{{H}^{2}}\,}, \qquad Z=\frac{{\rho }_{DE}}{3{{H}^{2}}}, \qquad \Delta =\frac{B}{3{{H}^{2}}\rho _{c}^{\alpha }},
\end{equation}
by making derivative the parameters $x$, $y$ and $Z$ with respect to $ln a$, and by using the continuity equations \eqref{conti2}, we can obtain them in the following form
\begin{subequations}\label{xyprime}
\begin{eqnarray}
&{x}'=3{{b}^{2}}x-3x-3Ax+3\Delta -2x\frac{{\dot{H}}}{{{H}^{2}}},\label{xyprime-1}\\
&{y}'=-3(1+{{\omega }_{m}})y-2\frac{{\dot{H}}}{{{H}^{2}}}y,\label{xyprime-2}\\
&{Z}'=-3{{b}^{2}}x-3Z-\frac{3}{2}\gamma (1-3{{\omega }_{m}})y+3Ax-3\Delta -2Z\frac{{\dot{H}}}{{{H}^{2}}},\label{xyprime-3}
\end{eqnarray}
\end{subequations}
where the prime denotes derivative with respect to $ln a$. We can find term $\frac{{\dot{H}}}{{H}^{2}}$ by Eq. \eqref{fri4-2} as
\begin{equation}\label{hdot2h2}
\frac{{\dot{H}}}{{{H}^{2}}}=-\frac{3}{2}-\frac{3}{2}{{\omega }_{m}}y-\frac{3}{4}\gamma y+\frac{9}{4}\gamma {{\omega }_{m}}y.
\end{equation}

Substituting \eqref{hdot2h2} into \eqref{xyprime}, the below relationships are deduced
\begin{subequations}\label{xyprime1}
\begin{eqnarray}
&{x}'=3{{b}^{2}}x-3Ax+3\Delta +3{{\omega }_{m}}xy+\frac{3}{2}\gamma xy-\frac{9}{2}\gamma {{\omega }_{m}}xy
,\label{xyprime1-1}\\
&{y}'=-3{{\omega }_{m}}y+3{{\omega }_{m}}{{y}^{2}}+\frac{3}{2}\gamma {{y}^{2}}-\frac{9}{2}\gamma {{\omega }_{m}}{{y}^{2}}
,\label{xyprime1-2}\\
&{Z}'=-3{{b}^{2}}x-\frac{3}{2}\gamma y+\frac{9}{2}\gamma {{\omega }_{m}}y+3Ax-3\Delta +3{{\omega }_{m}}yZ+\frac{3}{2}\gamma yZ-\frac{9}{2}\gamma {{\omega }_{m}}yZ.\label{xyprime1-3}
\end{eqnarray}
\end{subequations}

We note that the first Friedmann equation is rewritten by aforesaid new variables as
\begin{equation}\label{fried2}
  x+y+Z=1.
\end{equation}

Now we describe the stability conditions of the model to obtain the critical points that so-called fixed points. The corresponding fixed points have the asymptotical behavior, and they are dependent on the $f(T,\mathcal{T})$ model and modified Chaplygin gas parameters. In that case, we can obtain the fixed points in terms of the corresponding parameters by setting $x'=0$ and $y'=0$, and these results are demonstrated in Table \ref{table1}. As we can see, two fixed points are obtained by names $fp1$ and $fp2$. The significance of the fixed point is to study the stability of theory and phase transition.
\begin{table}[h]
\caption{The fixed points (Critical points).} 
\centering 
\begin{tabular}{|c |c |c |} 
\hline 
$Points$ & $fp1$ & $fp2$ \\ [0.5ex] 
\hline 
$x$ & \,\,\,\, $\frac{\Delta}{A-b^2}$\,\,\,\,\,\, & \,\,\,$\frac{\Delta}{A-b^2-\omega_m}$ \\
 \hline
$y$ &  0 & $\frac{2 \omega_m}{2 \omega_m+\gamma-3 \gamma \omega_m}$   \\
 [1ex] 
\hline 
\end{tabular}
\label{table1} 
\end{table}

In order to describe properties of the fixed points, we consider linear perturbations as $x' \rightarrow x' + \delta x'$ and $y' \rightarrow y' + \delta y'$ for Eqs. \eqref{xyprime1-1} and \eqref{xyprime1-2}. These perturbations give us two eigenvalues $\lambda_1$ and $\lambda_2$ for each of the fixed points. We note that stability condition occurs when the eigenvalues to be negative. In what follows, we find details of the eigenvalues of two critical points in table \ref{table2}.
\begin{table}[h]
\caption{Stability conditions of critical points.} 
\centering 
\begin{tabular}{|c| c| c|} 
\hline 
$Fixed\,\, points$ \,\,\,&\,\,\, $Eigenvalues$ \,\,\,&\,\,\, $Stability\,\, conditions$ \\ [0.5ex] 
\hline 
$fp1$ & \,\,\,\, $\lambda_1= -3 \omega_m$\,\,\,\,\,\, & $\omega_m > 0$ \\ 
& $\lambda_2= 3 b^2-3 A$ & $b^2 < A$\\
\hline
$fp2$ &  $\lambda_1= 3 b^2-3 A+3 \omega_m$ & $A > b^2+\omega_m$   \\
& $\lambda_2= 3 \omega_m$ &$\omega_m < 0$\\
 [1ex] 
\hline 
\end{tabular}
\label{table2} 
\end{table}

In order to investigate our model we need to describe the properties of fixed points so that the EoS parameter and the deceleration parameter are found as
\begin{subequations}\label{21}
\begin{eqnarray}
&\omega_{eff}=\frac{1}{2} \left(2 \omega_m +\gamma-3 \gamma \omega_m \right) y,\\
&q=-1-\frac{\dot{H}}{H^2}=\frac{1}{2}+\frac{3}{2}(2 \omega_m+\gamma -3 \gamma \omega_m) y.
\end{eqnarray}
\end{subequations}

Therefore, the corresponding results are seen in table \ref{table3}. We note that accelerated Universe occurs when $q < 0$. Tables \ref{table2} and \ref{table3} show us that condition $\omega_m < -\frac{1}{3}$ describes either accelerated Universe or stability condition.
\begin{table}[h]
\caption{Properties of the fixed points.} 
\centering 
\begin{tabular}{|c| c| c| c|} 
\hline 
$Fixed\,\, points$ \,\,\,&\,\,\, $\omega_{eff}$ \,\,\,&\,\,\,\,\, $q$  \,\,\,\,\, & \,\,\,\,\,$Acceleration$ \\ [0.5ex] 
\hline 
$fp1$ & \,\,\,\, $0$\,\,\,\,\,\, & $\frac{1}{2}$ & $No$\\
\hline 
$fp2$ &  $\omega_m$ & $\frac{1}{2}+\frac{3}{2} \omega_m$ &  $\omega_m < -\frac{1}{3}$  \\
 [1ex] 
\hline 
\end{tabular}
\label{table3} 
\end{table}

\section{Conclusion}\label{s5}
In this paper, we studied the $f(T,\mathcal{T})$ model by FRW background, in which $T$ and $\mathcal{T}$ were considered as torsion scalar and trace of the matter energy-momentum tensor, respectively. Then, we interacted the model with modified Chaplygin gas, this means that the effective energy density and the effective pressure of universe were written in terms of three components of matter, Chaplygin gas and dark energy. Since Universe considered as a perfect fluid, hence the effective continuity equation was separated by three continuity equations of Universe components. In that case, we supposed to exist only an energy flow between the components of dark energy and modified Chaplygin gas by interaction term as $Q= 3 b^2 H \rho_c$. In what follows, the arbitrary function $f(T,\mathcal{T})$ was considered as a linear function of $T$ and $\mathcal{T}$, and then Friedmann equation has been written in terms of redshift. Nonetheless, the dark energy cosmological parameters such as energy density, pressure and EoS parameter have drawn with respect to redshift. The corresponding figures showed that the Universe is undergoing accelerated expansion, and also EoS parameter is consistent with observational data.
Next, we investigated stability conditions for the model by using of phase plane analysis. Also, we obtained fixed points that these depended on parameters of the model and Chaplygin gas. We considered linear perturbations for the fixed points, and then we obtained the condition of accelerated expansion by calculating corresponding eigenvalues.


\end{document}